%% file: preprint.tex
\documentclass[11pt,a4paper,onecolumn]{article}
\usepackage[english]{babel}
\usepackage[utf8]{inputenc}
\usepackage{amsmath,amssymb}
\usepackage{graphicx}
\usepackage{dcolumn}
\usepackage{bm}
\usepackage[numbers,sort&compress]{natbib}

\usepackage[table]{xcolor} 
\usepackage{booktabs}
\usepackage{multirow}
\usepackage{threeparttable}

\usepackage{caption}
\usepackage{subcaption}

\usepackage{hyperref}

\usepackage[margin=1in]{geometry}

\begin{document}
	
\title{Extracting behavioural properties from face-to-face interactions temporal networks: a measure of egonet persistency}
	
\author{G. Maurial$^1$, E. Klüger$^2$, M. Génois$^1$}

\date{\small
		$^1$Aix Marseille Univ, Université de Toulon, CNRS, CPT, Marseille, France;\\
		$^2$Aix Marseille Univ, CNRS, LEST, Aix-en-Provence, France\\[2ex]
		\today
}

\date{\today}
\maketitle
\begin{abstract}
Understanding how individuals repeat social interactions over time is a central problem in the analysis of temporal networks. In social systems, repeated interactions shape processes such as information diffusion, collective coordination, and the emergence of social structure. Existing measures of egonet persistence often conflate genuine behavioural regularities with structural effects such as node degree, making it difficult to distinguish meaningful temporal correlations from random mixing.

In this work, we introduce the Neighbourhood Persistency Criterion (NPC), a statistically grounded framework for quantifying egonet persistence across time. NPC combines classical similarity measures with tailored null models controlling for network topology and interaction weights. 

We apply this framework to high temporal resolution face-to-face interaction networks collected at four Computational Social Science conferences using the SocioPatterns platform. Our results reveal a common behavioural structure across events, characterised by an exploration--exploitation trade-off in social interactions. While many individuals alternate between both strategies, others exhibit stable interaction patterns throughout the event. Importantly, these behaviours show little systematic association with socio-demographic attributes, suggesting that interaction strategies are shaped primarily by contextual factors rather than stable individual traits. NPC thus provides a flexible and interpretable tool for studying egonet persistence in temporal networks and social systems.
\end{abstract}

\vspace{10pt}

\section{\label{sec:intro}Introduction}

The study of temporal networks has emerged as a central framework for understanding dynamical social systems. Over the past two decades, the growing availability of high-resolution empirical data—derived from wearable sensors, Bluetooth devices, and digital platforms—has enabled researchers to observe human interaction patterns with unprecedented temporal precision. This empirical progress has fostered the development of temporal network theory at the intersection of statistical physics and computational social science~\cite{holme2012temporal, holme2015modern}. Pioneering initiatives such as the SocioPatterns collaboration have provided high-quality datasets capturing face-to-face interactions in a wide range of settings, including schools~\cite{stehle2011high, dai_longitudinal_2022}, hospitals~\cite{isella2011s}, workplaces, and scientific conferences~\cite{cattuto2010dynamics, barratSocialDynamicsConferences2010, genois_combining_2023}. These datasets record contacts at the scale of seconds or minutes, making it possible to move beyond static or time-aggregated representations of social networks and to investigate the fine-grained temporal organisation of human behaviour.

Building on these data, previous studies have identified several key temporal features of social networks, such as burstiness, memory effects, tie persistence, and circadian rhythms~\cite{masuda2016guide, karsai2012small}. A broad set of models and metrics has been introduced to characterise these properties, including inter-event time distributions~\cite{barabasi2005origin}, temporal motifs~\cite{kovanen2011temporal}, reachability and latency measures~\cite{pan2011path}, and temporal extensions of classical centrality indices~\cite{nicosia2013graph, takaguchi_coverage_2016}. These advances have enabled detailed investigations of dynamical processes on networks, such as epidemic spreading~\cite{holme2012temporal}, information diffusion, and influence propagation.

Despite this progress, understanding how individuals organise their social interactions over time remains challenging. In particular, we still lack interpretable and robust observables that directly quantify the \emph{persistence of interactions at the egocentric level}, a key ingredient for understanding socially meaningful phenomena such as coordination, cohesion, and repeated social engagement. 

This issue is particularly salient in the context of academic conferences, where the global repartition of interactions is quite heterogeneous compare to other social gathering places \cite{colmanSocialFluidityMobilizes2021}. Conferences offer valuable networking opportunities while also enabling geographically distant collaborators to meet face-to-face. Given the limited duration of these events, the large number of attendees, and the abundance of potential future collaborators, participants must continuously balance the maintenance of existing ties with opportunities for exploration. This setting raises a central empirical question: to what extent do individuals repeatedly interact with the same partners across social events, and how stable are these egocentric interaction patterns over the course of a conference?

To investigate this question, we analyse high temporal resolution face-to-face interaction networks collected at several Computational Social Science (CSS) conferences~\cite{genois_combining_2023}. To support this analysis, a family of egonet-based temporal observables, termed the \textbf{Neighbourhood Persistency Criterion (NPC)}, which quantify the persistence and recurrence of interactions within an individual’s egocentric network  is introduced in section \ref{sec:method}. 

Using this framework, we show in Section~\ref{sec:appli}, that conference attendees exhibit robust and recurrent behavioural patterns that are remarkably consistent across events. In particular, we identify a characteristic exploration–exploitation structure in social interactions: while many individuals alternate between both strategies across social breaks, others display stable interaction patterns throughout the event. Importantly, we find that these behavioural dynamics are only weakly associated with socio-demographic or self-reported individual traits, suggesting that interaction strategies are shaped primarily by contextual and situational factors rather than stable personal characteristics. Finally, Section~\ref{sec:discu} discusses the implications and limitations of these findings and outlines open questions for future work.

\section{\label{sec:method}Method}

\subsection{Setting Grounded Similarity Measures for Egonet Persistence}

A temporal network can be represented at each time $t$ by an adjacency matrix ${A^t}$, where $A^t_{ij} = 1$ if nodes $i$ and $j$ are connected at time $t$, and $A^t_{ij} = 0$ otherwise. The aggregated network over a period $T$ is then defined as $A^T$, such that $A^T_{ij} = \sum\limits_{t \in T} A^t_{ij}$. The set of neighbours of node $i$ during period $T$ is given by $V^T_i = \{j \,|\, A^T_{ij} > 0 \}$.

Inspired by previous studies on the persistence of neighbourhood structures in temporal networks~\cite{valdano_predicting_2015, escribanoStabilityPersonalRelationship2023}, a natural first step is to compare a node’s egocentric network across two distinct time periods, $T$ and $T'$, using similarity measures. To this end, we compute both the Jaccard index and cosine similarity for each node $i$:

\begin{equation}
	\lambda_J^{T,{T'}}(i) = \dfrac{|V^T_i \cap V^{T'}_i|}{|V^T_i \cup V^{T'}_i|} \in [0,1],
	\label{eq:def}
\end{equation}

\begin{equation}
	\lambda_{\mathrm{cos}}^{T,{T'}}(i) = \dfrac{\sum\limits_{j \in V^T_i \cup V^{T'}_i}  A^T_{ij} \, A^{T'}_{ij}}{\sqrt{\sum\limits_{j \in V^T_i}  (A^T_{ij})^2 \, \sum\limits_{j \in V^{T'}_i}  (A^{T'}_{ij})^2}} \in [0,1].
	\label{eq:def2}
\end{equation}

\begin{figure}[t]
	\centering
	\begin{tabular}{cc}
		\begin{subfigure}[c]{0.47\textwidth}
			\centering
			\includegraphics[width=\textwidth]{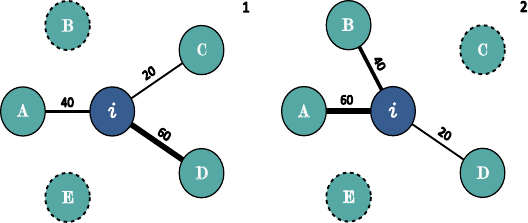}
		\end{subfigure} &
		\begin{subfigure}[c]{0.47\textwidth}
			\centering
			\includegraphics[width=\textwidth]{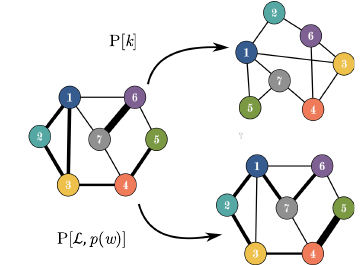}
		\end{subfigure} \\
		
		\begin{subfigure}[t]{0.47\textwidth}
			\caption{Weighted egonet representation of node $i$ at two distinct periods. In this example, $\lambda^{1,2}_J(i) = 1/2$ and $\lambda^{1,2}_{\mathrm{cos}}(i) = 5/7$, recovering the fact that node $i$ maintains more high-weights interactions.}
			\label{fig:defA}
		\end{subfigure} &
		\begin{subfigure}[t]{0.47\textwidth}
			\caption{Illustration of the randomisation effect on the weighted network. $P[k]$ preserves the degree of each node, rewiring links, giving back an unweighted network, while $P[\mathcal{L}, p(w)]$ preserves the network topology but randomises link weights according to the original weight distribution}
			\label{fig:defB}
		\end{subfigure}
	\end{tabular}
	
	\caption{Representation of the methodology used for the framework.}
	\label{fig:definition}
\end{figure}

\noindent
$A^T_{ij}$ denotes the total interaction duration between nodes $i$ and $j$ during period $T$, corresponding to the weight of edge $i$–$j$. 

The two observables serve complementary roles in quantifying the temporal stability of an individual's egocentric network. The Jaccard index, $\lambda_J^{T,{T'}}(i)$, measures the overlap in contact identities between two periods, independent of interaction strength, making it particularly useful for assessing partner retention.The cosine similarity, $\lambda_{\mathrm{cos}}^{T,{T'}}(i)$, incorporates interaction weights, thereby capturing not only whether contacts are renewed but also whether their intensity is preserved. This distinction is particularly relevant in social contexts, where repeated short encounters may differ qualitatively from fewer but longer exchanges. The combined use of both metrics thus provides a more nuanced view of egonet persistence. For example, in Fig.~\ref{fig:definition}.(a), node $i$ maintains more interactions with high-weight contacts in the second period, resulting in a higher cosine similarity value ($\lambda^{1,2}_{\mathrm{cos}}(i) = 5/7$) compared to the Jaccard similarity ($\lambda^{1,2}_J(i) = 1/2$).

Raw similarity scores alone are insufficient to capture meaningful temporal correlations in social behaviour, as both measures are inherently sensitive to node degree. Indeed, individuals who consistently interact with a large portion of the population (i.e., with a degree exceeding half the total number of nodes) will exhibit artificially high similarity values, independent of true social preferences or behavioural persistence.

To address this bias, we introduce a statistical normalisation framework based on randomisation. Specifically, we apply two null models to each time period, illustrated in Fig.~\ref{fig:definition}(b): $P[k]$, which rewires the network while preserving node degrees, and $P[\mathcal{L}, p(w)]$, which retains the network topology but randomly redistributes interaction weights according to the empirical distribution~\cite{gauvin_randomized_2022}.

By rewiring the network, $P[k]$ allows us to control for the correlation between the observable and network topology. Similarly, $P[\mathcal{L}, p(w)]$ eliminates correlations between topology and interaction intensity, enabling the study of repeated interaction strengths. From this characteristics, $P[k]$ will be considered the null model for Jaccard similarity, while $P[\mathcal{L}, p(w)]$ concerns cosine similarities.

Within this framework, we can compute standardised similarity scores, such as $z$-scores, to quantify the deviation of an observed pattern from what would be expected under null conditions. Building on this idea, we define a revised observable, \textbf{Neihbourhood Persistency Criterion (NPC)}, based on the median of node $i$'s null distribution, $\mu_{l}(i)$, as follows:

\begin{equation}
	\alpha_l(i) =
	\begin{cases}
		\dfrac{1}{1-\mu_{l}(i)} \int_{\mu_{l}(i)}^{\lambda_l(i)}{\rho_l(x)dx}, & \text{if } \lambda_l(i) \geq \mu_{l}(i) \\[10pt]
		\dfrac{1}{\mu_{l}(i)} \int_{\lambda_l(i)}^{\mu_{l}(i)}{\rho_l(x)dx}, & \text{if } \lambda_l(i) \leq \mu_{l}(i)
	\end{cases}
	\, \in [-1, 1], \, l \in \{J, \mathrm{cos}\}
	\label{eq:signed_p}
\end{equation}

\noindent
Here, $\rho_l$ denotes the density function of the null distribution for either the Jaccard or cosine similarity measure. As for a one-tailed $p$-value, this observable can be interpreted as the probability, under the null hypothesis, of drawing a value between the empirical observation and the median of the null distribution—conditional on the draw occurring on the same side of the median. In contrast to the one-tail p-value, this formulation allows us to distinguish between positively and negatively deviating behaviours with continuous values while preserving a probabilistic interpretability of the measure.

Omitting node and similarity specification, $\alpha$ quantifies the extent to which the observed egonet similarity deviates from the randomised expectation, accounting for both skewness and variability in the null distribution. In particular, $\alpha = -1$ indicates a strong avoidance of repeated interactions, while $\alpha = 1$ reflects a strong preference for re-engaging with the same individuals. Values near $\alpha = 0$ indicate no deviation from random interaction patterns, consistent with null expectations.

In summary, $\alpha_J$ and its weighted counterpart $\alpha_{\mathrm{cos}}$ provide behaviourally grounded, degree-normalised measures of egonet persistence. Having defined these observables, the next step is to examine how they evolve over time for each individual and how different behavioural patterns emerge across the population. This requires moving from pointwise comparisons between periods to a higher-level view of the full temporal trajectory of each node.

\subsection{Dimension Reduction}

To characterise and categorise interaction dynamics across the dataset, we represent each node by a vector of its $\alpha$ values computed over all pairs of defined time periods. In this representation, behavioural similarity corresponds to proximity in a high-dimensional space. Classical clustering or dimensionality-reduction techniques—such as Principal Component Analysis (PCA), $k$-means, or DBSCAN—could, in principle, be applied directly to this space to identify common behavioural types.

However, such global clustering approaches may obscure meaningful patterns, since behavioural similarity may arise only for specific combinations of time periods, without forming consistent behavioural groups overall. Moreover, there is no a priori reason to assume that individuals maintain a stable behavioural type across all time periods. A typical way to reduce dimensionality would be to consider each individual's mean $\alpha$ value (denoted $\overline{\alpha}$), thereby capturing their overall behaviour across time. Yet this approach is insufficient: for instance, an individual alternating between engagement and avoidance across different periods could end up with a $\overline{\alpha}$ close to zero, masking the underlying variability of their behaviour.

This motivates an alternative strategy: focusing on the temporal consistency of each individual's behaviour through both the mean and a measure of dispersion. Since there is no reason to expect the distribution of $\alpha$ values to be normal, the standard deviation may not constitute an appropriate dispersion estimator. It is therefore necessary to compare different dispersion estimators in order to identify the most suitable one.

We employ the NPC framework as the basis for behavioural categorisation because it enables the definition of clear-cut behavioural archetypes, such as vectors with only  $-1$, $0$, or $1$ coordinates, which represent, respectively, repeated avoidance of previous encounters, constant random exploration of the sample pool, and strong re-engagement. These archetypes are essential for selecting an appropriate dispersion estimator. The archetypes, together with the detailed methodology used to identify the most suitable estimator, are presented in the Supplementary Information~\ref{sec:dispersion}.

Standard deviation performed poorly in capturing behavioural variability in this setting, due to its sensitivity to outliers. After benchmarking several archetype alternatives, we found that the Trimmed Mean Dispersion using Tukey’s method (TMDT), provided the most robust and interpretable measure of intra-individual behavioural consistency.

TMDT relies on trimming, depending on a percentile $q$, of the lowest and highest values in each vector before computing dispersion, as follows. Let's consider a vector $X$ of scalars. If we denote by $Q_q(X)$ the q-th percentile of $X$ then it's inter q-th percentile range (IPR) is denoted by $IQR_q(X) = Q_{100-q}(X) - Q_q(X)$. The main idea from Tukey's was to consider outliers to be values outside the window $\mathbb{W}(X,q)=[Q_{q}(X)-1.5 IQR_q(X) , ~ Q_{100-q}(X)+ 1.5 IQR_q(X)]$ (with $q = 25$ usually). Thus, writing $\overline{X}$ the average value of $X$, TMDT is defined as:

\begin{equation}
	\text{TMDT}_q(X) = \dfrac{\sum\limits_{x_i \in X} |x_i -\overline{X}| \mathbb{I}_{\mathbb{W}(X,q)}(x_i)}{\sum\limits_{x_i \in X} \mathbb{I}_{\mathbb{W}(X,q)}(x_i)} 
	\label{eq:TMDT}
\end{equation}
\noindent
with $\mathbb{I}_{\mathbb{W}(X,q)}$ the indicator function for the interval $\mathbb{W}(X,q)$.

Across all conference datasets, we found extremely similar values for a range of $q$ between $5\,\%$ and $20\,\%$, with the most stable values around $q = 12 \, \%$ (see Supplementary Information~\ref{sec:rob_q}).

Based on these results, we use TMDT$_{12}$ (i.e. $q = 12\, \%$) throughout the analysis as a reliable estimator of egonet persistence consistency over time. Having established NPC and a robust measure of its temporal variability, we now turn to their application on empirical data. This transition from methodological development to real-world datasets allows us to assess how the proposed framework behaves under the complex, bursty dynamics of actual human interactions.

\section{\label{sec:appli}Application to conferences}

We applied the above methodology to face-to-face interaction networks collected at four conferences in computational social science: the 3rd GESIS Computational Social Science Winter Symposium (WS16), the International Conference on Computational Social Science (ICCSS17), the Eurosymposium on Computational Social Science (ECSS18) and the 41st European Conference on Information Retrieval (ECIR19)~\cite{genois_combining_2023}. These datasets were obtained using wearable proximity sensors with a temporal resolution of 20 seconds. While this high resolution captures fine-grained interaction dynamics, it also reflects the inherently bursty nature of human activity at short timescales~\cite{barabasi2005origin}. Consequently, applying NPC directly at each time step would not yield meaningful insights.

To address this, we first identified the relevant temporal windows over which to aggregate interactions. Specifically, we analysed the activity timeline of each network and extracted contiguous periods of high interaction intensity. These periods correspond to socially meaningful events within the conference schedule—coffee breaks, lunch, and poster sessions—during which attendees are free to move and interact \cite{genois_combining_2023}. Interactions within each of these high-activity windows were aggregated to construct a sequence of weighted contact networks, providing an appropriate temporal granularity for studying the persistence and recurrence of social behaviour using NPC framework. Individuals with recorded interaction in less than two breaks are treated as non-attending, corresponding to 1~\% at 5~\% of active nodes over all datasets (see Supplementary~\ref{sec:period} for global activity timeline and Supplementary~\ref{sec:nodes} for exact values).

In addition, we removed all contacts with a duration of exactly 20 seconds to reduce spurious cross-path detections caused by high local density. To assess the robustness of our results to interaction-duration filtering, we recomputed both $\alpha$ after excluding contacts shorter than 0, 20, 40, 60, 120, and 300 seconds. We found that the resulting measures are highly sensitive to such filtering choices. Therefore, the selection of a threshold and the filtering method must be carefully justified by the researcher according to the study context, especially knowing that short contact removal is not always the best option \cite{elmerValidityRFIDBadges2019}.
A description of the robustness under filtering and the characterisation of each temporal windows is provided in the Supplementary Informations~\ref{sec:filter}. Finally, each randomisation presented in Section~\ref{sec:method} was performed a thousand time for each period.

\subsection{Typical conference behaviour}

\begin{figure}
	\centering{\includegraphics[width=1\textwidth]{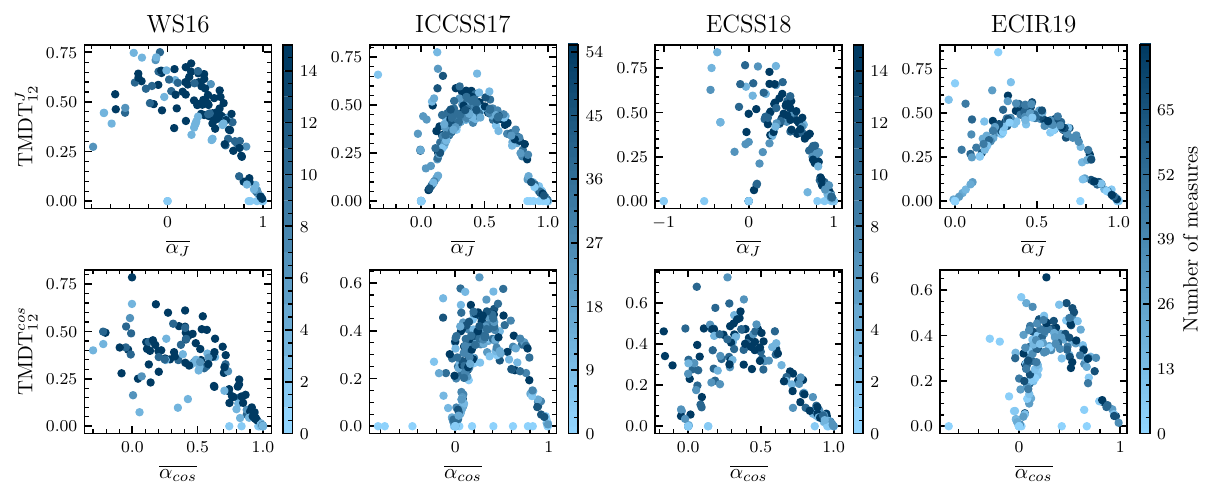}}
	\vspace*{8pt}
	\caption{TMDT$_{12}$ versus $\overline{\alpha}$ for each node in each conference. For nodes with high dispersion values, $\overline{\alpha}$ is not reliable, whereas for consistent nodes they are meaningful. Point colours indicate the number of measures available for each node (i.e. the number of couple of periods where the node was active) across the full dataset.}
	\label{fig:clustering}
\end{figure}

Examining $\text{TMDT}_{12}$ against $\overline{\alpha}$ for each similarity measure (Fig.~\ref{fig:clustering}), we find a reversed U-shaped relationship across all conferences, suggesting a shared global behavioural pattern (for robustness over contact duration filtering see Supplementary Informations~\ref{sec:rob_filt}). Most attendees have $\overline{\alpha}$ that are either positive or close to zero, indicating that strong avoidance behaviours are rare.

This joint measure naturally separates attendees into four broad behavioural types. Individuals with low dispersion show consistent interaction patterns across social breaks. Within this group, some maintain a stable set of contacts over time ($\overline{\alpha}$ close to 1), while others frequently change partners without a clear structure ($\overline{\alpha}$ near zero), possibly reflecting an intent to network widely as reported on a previous conference study \cite{kordts-freudingerLearningInteractionConference2017}. Finally, a small
group with large negative values stands out: they are characterised by both low attendance across social breaks and a systematic avoidance of previous contacts ($\overline{\alpha}$ close to -1, light blue points).

In contrast, individuals with high dispersion do not have a clear pattern during social breaks. We expect that the alternation between exploration and exploitation strategies is producing such large variability in their $\alpha$ values which cannot be captured by the average alone.

One possible explanation for this behavioural switching is a gradual selection process in which attendees progressively narrow their social circles over time, keeping only their most relevant contacts. If this were the case, we would expect a systematic increase in egonet similarity across consecutive periods pairs. However, our analysis of $\alpha$ distributions for consecutive period pairs shows no such pattern for $\alpha_{J}$, and even the opposite trend for $\alpha_{\mathrm{cos}}$ in the WS16 and ECSS18 datasets (see Supplementary~\ref{sec:dyna}). Moreover, egonet persistence does not significantly depend on the temporal distance between periods.

This suggests that the observed alternation between exploration and exploitation is not a by-product of gradual social filtering, but rather a distinct and recurring behavioural mode.

\subsection{More insight from NPC comparison}

\begin{figure}
	\centering{\includegraphics[width=1\textwidth]{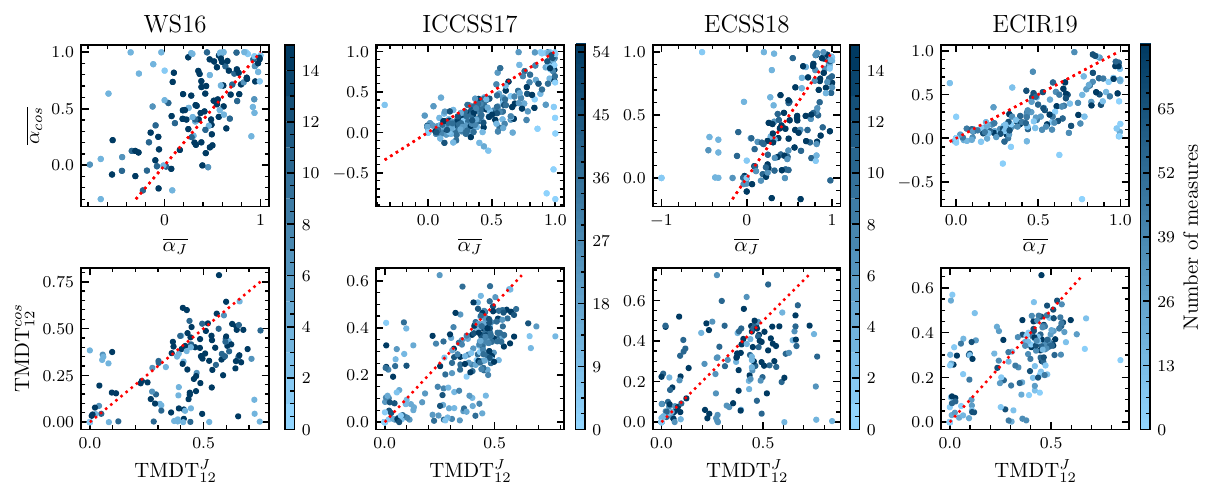}}
	\vspace*{8pt}
	\caption{Comparison of both $\overline{\alpha}$ (top line) and TMDT$_{12}$ (bottom line) for each node in each conference. The red dotted line corresponds to the curve $y = x$. The colour of each point indicates the number of measures available for that node in the full dataset, i.e. the number of couple of periods in which the node was active.}
	\label{fig:corr_NPC}
\end{figure}

Even though similar patterns emerge for both NPC observables, this does not imply that each individual exhibits comparable values for \emph{both}. As shown in Fig.~\ref{fig:corr_NPC} (top line), $\overline{\alpha_J}$ and $\overline{\alpha_{\mathrm{cos}}}$, form a cloud of points lying close to the curve $y=x$ for all conferences, indicating a strong correlation between the two persistence measures. Except for WS16, the point clouds lie predominantly below the $y=x$ line, suggesting that attendees, on average, tend to be slightly more persistent in their choice of contacts than in the duration of those interactions. In other words, maintaining the same neighbourhood across periods appears easier than allocating similar amounts of time to each neighbour from one period to the next.

When considering behavioural consistency (bottom line in \ref{fig:corr_NPC}), attendees can be grouped into three categories:  
(i) those who behave consistently (or switch values) in $\alpha_{J}$ and $\alpha_{\mathrm{cos}}$, lying close to the red curve;  
(ii) those who are consistent in $\alpha_{J}$ only, staying at the bottom right corner of the plots;  
(iii) those who are consistent in $\alpha_{\mathrm{cos}}$ only, closer to the top left corner of the panels.  
Latter two categories ((ii) and (iii)) are particularly surprising, as one might expect that patterns of exploration or exploitation would produce similar levels of consistency in both contact identity and interaction duration.

We currently lack sufficient information to give a definitive interpretation of categories (ii) and (iii), but we can propose some plausible scenarios. For category (ii), two distinct cases seem possible. First, the attendee may display strong persistence in $\alpha_{J}$, but distribute their interaction time differently depending on the type of break (e.g., always engaging with one group at lunch and another at coffee breaks, while still encountering members of both groups). Second, $\overline{\alpha_J}$ may remain close to zero, with the inconsistency in $\alpha_{\mathrm{cos}}$ arising from recurrent, intense interactions with a small subset of contacts (e.g., lengthy conversations with the same few people during certain breaks). For category (iii), two scenarios are also conceivable. In one, $\overline{\alpha_{\mathrm{cos}}}$ remains close to 1, indicating that the attendee consistently spends long periods with the core of their social circle across all breaks, regardless of changes in contact identity. In the other, the attendee adopts an exploratory strategy in terms of time allocation—varying interaction durations considerably—while such variation has little bearing on whether their contact set is stable or shifting.

To summarise, these interpretations highlight that consistency in one measure but not the other may reflect qualitatively different interaction strategies. Understanding whether such patterns are systematically related to individual characteristics—such as socio-demographic attributes, personality traits, or motivational profiles—requires a closer examination of metadata. 

\subsection{Metadata correlation}

\begin{table}[h]
\centering
\begin{threeparttable}
	\caption{Significant sociodemographic–NPC correlations identified using the Brunner–Munzel and Kolmogorov–Smirnov tests.}
	\label{tab:correl}
	{\begin{tabular}{l|l|cc|cc|cc|cc}
			\toprule
			\multicolumn{2}{c|}{} & \multicolumn{2}{c|}{WS16} & \multicolumn{2}{c|}{ICCSS17} & \multicolumn{2}{c|}{ECSS18} & \multicolumn{2}{c}{ECIR19} \\
			\multicolumn{2}{c|}{} & $\alpha_{J}$ & $\alpha_{\mathrm{cos}}$ & $\alpha_{J}$ & $\alpha_{\mathrm{cos}}$ & $\alpha_{J}$ & $\alpha_{\mathrm{cos}}$ & $\alpha_{J}$ & $\alpha_{\mathrm{cos}}$ \\
			\midrule
			
			Age & below 30 & X & X & (+) & (+) & (+) & (+) & (+) & X \\
			& 30–39 & X & X & X & (–) & X & X & X & X \\
			& over 40 & X & (–) & X & (+) & X & X & X & X \\
			
			\midrule
			Gender & Male & X & X & X & X & X & X & X & X \\
			& Female & X & X & X & X & X & X & (+) & (+) \\
			
			\midrule
			Language & 1 & X & (–) & (+) & (+) & (+) & X & X & X \\
			& 2 & (+) & X & (–) & (–) & X & X & (–) & (–) \\
			& 3 & . & . & (+) & (+) & . & . & X & X \\
			& 4 & . & . & X & X & . & . & . & . \\
			& 5 & . & . & X & X & . & . & . & . \\
			& 6 & . & . & X & (–) & . & . & . & . \\
			
			\midrule
			Known Persons & 0 & . &  & (–) & (–) & X & X & (–) & (–) \\
			& 1–5 & . & . & X & X & X & X & (+) & (+) \\
			& 6–11 & . & . & X & X & X & X & (+) & X \\
			& 11–20 & . & . & (+) & X & X & X & (–) & (–) \\
			& over 20 & . &.  & (+) & (+) & X & X & X & X \\
			
			\bottomrule
	\end{tabular}}
	\begin{tablenotes}[flushleft]
	\small
	\item Only subgroups with more than 10 individuals were included. A blank cell indicates missing metadata, while ``X'' marks the absence of significant correlations. ``(+)'' (respectively ``(–)'') denotes that the subgroup exhibits higher (respectively lower) overall $\alpha$ values than the full observed population. For anonymisation, languages are represented by numbers ranked from the most to the least frequent; these do not correspond to the same languages across conferences.
	\end{tablenotes}

\end{threeparttable}
\end{table}

To investigate whether the behavioural categories described above are systematically linked to individual characteristics, we next compared the distributions of TMDT$_{12}$ values across socio-demographic variables, such as age, gender, academic status, and specific traits as the role at conference or number of known attendees, useful for assessing major situational characteristics (all metadata collected are described in~\cite{genois_combining_2023}). Given the fuzzy boundaries between clouds of points in Fig.~\ref{fig:clustering}, we opted for distributional rather than categorical correlation testing. While it would be possible to impose thresholds for a categorical analysis, such thresholds must be chosen with care, as significance levels are highly sensitive to these choices~\cite{lylesSensitivityAnalysisMisclassification2010}.

Because the TMDT$_{12}$ distributions deviate from normality and may violate equal-variance assumptions, parametric tests such as the Student’s $t$-test were deemed inappropriate. Instead, we employed the Brunner–Munzel test~\cite{brunnerNonparametricBehrensFisherProblem2000, karchBmtestJamoviModule2023} and the Kolmogorov–Smirnov test~\cite{EDFStatisticsGoodnessstephens1974}, both of which are robust to these violations. A correlation was considered significant only if at least one of the two tests yielded a $p$-value below $0.001$ and the other below $0.01$. We applied the same procedure to the full distribution of $\alpha$ values. While the majority of significant results are reported in the Supplementary Informations~\ref{sec:metadata_corr}, a subset of results for the most basic socio-demographic traits is presented in Table~\ref{tab:correl}.

Across all conferences and variables tested, no significant associations emerged for the behavioural consistency categories. This lack of correlation suggests that typical interaction strategies—whether consistent or variable—are not strongly determined by stable individual traits or self-reported expectations. Instead, these behaviours appear to be shaped by contextual factors such as event structure, scheduling constraints, or individual objectives during the conference.

Nonetheless, some weak trends hint at context-dependent strategies rather than stable individual predispositions. For instance, in \textit{Age} category, attendees younger than 30 years tend to exhibit more persistent interactions than the overall population in three of the four conferences. By contrast, senior attendees display heterogeneous behaviours across conferences, showing higher $\alpha_{\mathrm{cos}}$ in ICCSS17 but lower ones in WS16. This age-related effect may reflect contextual factors, such as the presence of small groups of Master’s students who are less invested in networking compared to late-stage PhD students actively seeking postdoctoral opportunities. Overall, these findings reinforce the view that common behavioural patterns can emerge within specific groups at a given event without being tightly coupled to stable individual traits.

\section{\label{sec:discu}Discussion}

In this paper, we introduced \textbf{NPC}, an analytical tool for quantifying egonet persistency in temporal networks. Applying NPC to face-to-face interaction data collected at academic conferences, we demonstrated their ability to differentiate behavioural dynamics, thereby providing a novel tool for the behavioural description of interaction patterns in social gatherings.

Although the primary focus of this work was the formal definition and validation of NPC, our analyses also yielded substantive insights into interaction dynamics at conferences. We found that attendees do not avoid previous contact partners during breaks; instead, their behaviour reflects an \emph{exploration–exploitation dilemma}. While many participants switch between exploratory and exploitative strategies from one break to another, some remain consistent in their approach throughout the event. Importantly, our correlation analyses indicate that these behavioural tendencies are not strongly associated with stable individual traits, suggesting they are shaped primarily by contextual factors such as personal timing, individual objectives, and situational constraints.

There are several promising directions for future research. In this study, we only partially examined the temporal evolution of persistency, focusing on aggregated measures over the entire conference. We did not investigate daily-level persistency patterns or perform cross sectional analysis. These additional perspectives could yield a more nuanced description of individual behaviour in conferences or other social contexts.

The examples presented here represent only a subset of the phenomena that persistency studies can reveal. A promising direction would be to extend this framework to investigate link-level persistency, with the aim of capturing even finer-grained aspects of temporal interaction structure. As empirical temporal network datasets continue to grow in availability and diversity, we expect NPC to become a valuable analytical tool for studying social dynamics across a wide range of environments.

\section*{Acknowledgments}
	This work was supported by ENS Paris-Saclay. We thank the Centre de Physique Théorique, Aix-Marseille University, and the CNRS for hosting, and the CPT network team for their valuable insights, discussions, and contributions.

\bibliographystyle{abbrvnat}
\bibliography{main}

\newpage
\section*{Supplementary Informations}
\appendix
\input{SM.tex}

\end{document}

%% file: SM.tex
\section{\label{sec:dispersion}Dispersion estimators}

A wide range of dispersion estimators exists; however, most studies rely almost exclusively on the conventional standard deviation rather than adopting measures tailored to the data at hand. In this part, we introduce several alternative dispersion estimators and benchmark their performance in our specific context.

\subsection{Definitions}

\begin{figure}[h!]
	\centering
	\includegraphics[width=1\textwidth]{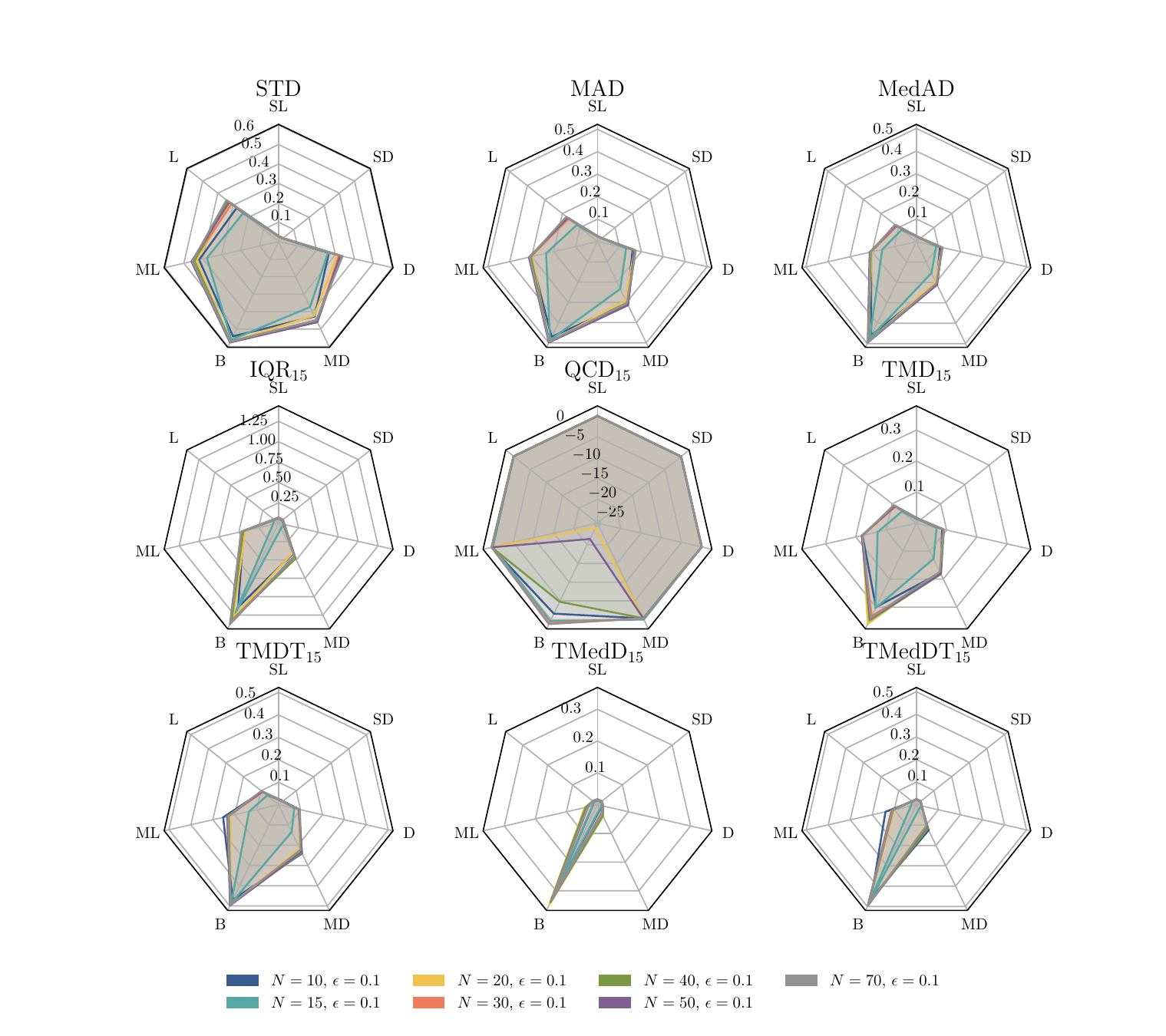}
	\caption{Radar plots of the resulting values for each dispersion estimators applied for all archetypes following the fixed percentage archetype's definition, with noise $\epsilon = 0.1$ and sample size $N$ between 10 and 70.}
	\label{fig:all_esti}
\end{figure}

Let $X = \{x_1,\dots,x_N\}$ be a vector of $N$ scalars with mean value $\overline{X}$ and median $\tilde{X}$.  
We introduce the following dispersion estimators:

\begin{equation}
	\mathrm{STD}(X) = \sqrt{\dfrac{1}{N} \sum\limits_{x \in X} (x - \overline{X})^2}
\end{equation}

\begin{equation}
	\mathrm{MAD}(X) = \dfrac{1}{N} \sum\limits_{x \in X} |x - \overline{X}|
\end{equation}

\begin{equation}
	\mathrm{MedAD}(X) = \dfrac{1}{N} \sum\limits_{x \in X} |x - \tilde{X}|
\end{equation}

If we denote by $Q_q(X)$ the $q$-th percentile of $X$, then its inter-$q$th percentile range (IQR) is given by:

\begin{equation}
	\mathrm{IQR}_q(X) = Q_{100-q}(X) - Q_q(X)
\end{equation}

And the Quartile Coefficient of Dispersion (QCD) is defined by:

\begin{equation}
	\mathrm{QCD}_q(X) = \dfrac{Q_{100-q}(X)-Q_q(X)}{Q_{100-q}(X)+Q_q(X)}
\end{equation}

Because some values may be outliers, we also consider trimmed dispersion estimators. Let $\mathbb{V}(X,q) = [Q_{q}(X), Q_{100-q}(X)]$ denote the basic trimming window. Then the trimmed mean and median absolute deviations are:

\begin{equation}
	\mathrm{TMD}_q(X) =
	\dfrac{\sum\limits_{x_i \in X} |x_i -\overline{X}|\,
		\mathbb{I}_{\mathbb{V}(X,q)}(x_i)}
	{\sum\limits_{x_i \in X} \mathbb{I}_{\mathbb{V}(X,q)}(x_i)}
\end{equation}

\begin{equation}
	\mathrm{TMedD}_q(X) =
	\dfrac{\sum\limits_{x_i \in X} |x_i -\tilde{X}|\,
		\mathbb{I}_{\mathbb{V}(X,q)}(x_i)}
	{\sum\limits_{x_i \in X} \mathbb{I}_{\mathbb{V}(X,q)}(x_i)}
\end{equation}

This direct trimming is often considered too strict. Tukey’s method instead defines a larger window to exclude only extreme outliers:  
\[
\mathbb{W}(X,q) =
\bigl[\,Q_{q}(X)-1.5\,\mathrm{IQR}_q(X),~
Q_{100-q}(X)+1.5\,\mathrm{IQR}_q(X)\,\bigr],
\]
with $q=25$ by default. Using this window, the Tukey-trimmed mean and median absolute deviations become:

\begin{equation}
	\mathrm{TMDT}_q(X) =
	\dfrac{\sum\limits_{x_i \in X} |x_i -\overline{X}|\,
		\mathbb{I}_{\mathbb{W}(X,q)}(x_i)}
	{\sum\limits_{x_i \in X} \mathbb{I}_{\mathbb{W}(X,q)}(x_i)}
\end{equation}

\begin{equation}
	\mathrm{TMedDT}_q(X) =
	\dfrac{\sum\limits_{x_i \in X} |x_i -\tilde{X}|\,
		\mathbb{I}_{\mathbb{W}(X,q)}(x_i)}
	{\sum\limits_{x_i \in X} \mathbb{I}_{\mathbb{W}(X,q)}(x_i)}
\end{equation}

We compared these nine dispersion estimators on a set of predetermined archetypes designed to represent distinct behavioural profiles. We defined seven types of behaviours:  
\begin{enumerate}
	\item \textbf{Super Loyal (SL)}: all $\alpha$ values close to $1$.
	\item \textbf{Loyal (L)}: most $\alpha$ values close to $1$, with a few close to $0$.
	\item \textbf{Mildly Loyal (ML)}: a considerable number of $\alpha$ values close to $1$, the remainder close to $0$.
	\item \textbf{Bursty (B)}: values uniformly distributed between $-1$ and $1$.
	\item \textbf{Mildly Disloyal (MD)}: a considerable number of $\alpha$ values close to $-1$, the remainder close to $0$.
	\item \textbf{Disloyal (D)}: most $\alpha$ values close to $-1$, with a few close to $0$.
	\item \textbf{Super Disloyal (SD)}: all $\alpha$ values close to $-1$.
\end{enumerate}

\subsection{Chosen estimator}

\begin{figure}[h!]
	\centering
	\includegraphics[width=1\textwidth]{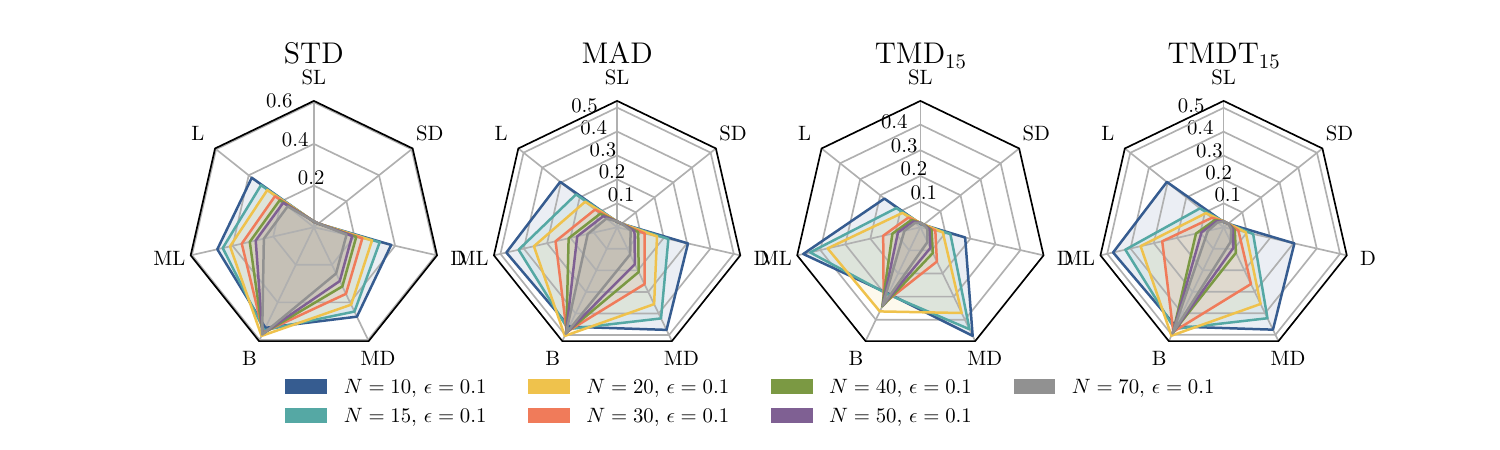}
	\caption{Radar plots of the resulting values for best dispersion estimators applied for all archetypes following the second archetype's definition with noise $\epsilon = 0.1$ and sample size $N$ between 10 and 70.}
	\label{fig:good_esti}
\end{figure}

Numerically, the notion of “closeness” was implemented by adding random noise $\sigma$ uniformly distributed over $[-\epsilon,\epsilon]$. For values close to $1$ (resp.\ $-1$), we subtracted $|\sigma|$ from $1$ (resp.\ added $|\sigma|$ to $-1$), whereas for values close to $0$ we added $\sigma$.  

We constructed two types of synthetic samples for different sample sizes $N$, covering the range observed in the conference datasets. In the first, we inserted a fixed number of values close to $0$ (2 for L and D, 5 for ML and MD), and in the second we inserted a fixed \emph{percentage} of values close to $0$ (10\% for L and D, 20\% for ML and MD).These values were selected randomly.

Figure~\ref{fig:all_esti} shows the estimators applied to the fixed-percentage archetypes after averaging over 100 trials. All estimators correctly reflect that the Bursty (B) type has the highest dispersion.  However, QCD, IQR, TMedD and TMedDT provide poor discrimination between the SL, L and ML archetypes (and symmetrically between SD, D and MD). We therefore eliminated these estimators from further analysis. We also excluded MedAD, as our comparisons are based on mean values rather than medians.

The second benchmarking experiment, shown in Figure~\ref{fig:good_esti}, uses the fixed-number archetypes. We observe that TMD performs poorly for the Bursty (B) type at small $N$, whereas the remaining estimators give broadly consistent results. TMDT exhibits the most interesting behaviour: its variation increases with $N$, reflecting the higher probability that values close to $0$ are classified as outliers for all archetypes except Bursty. This property makes TMDT the most informative and robust estimator for our context.

\newpage

\section{\label{sec:rob_q}Robustness under $q$}

\begin{figure}[h!]
	\centering
	\includegraphics[width=1\textwidth]{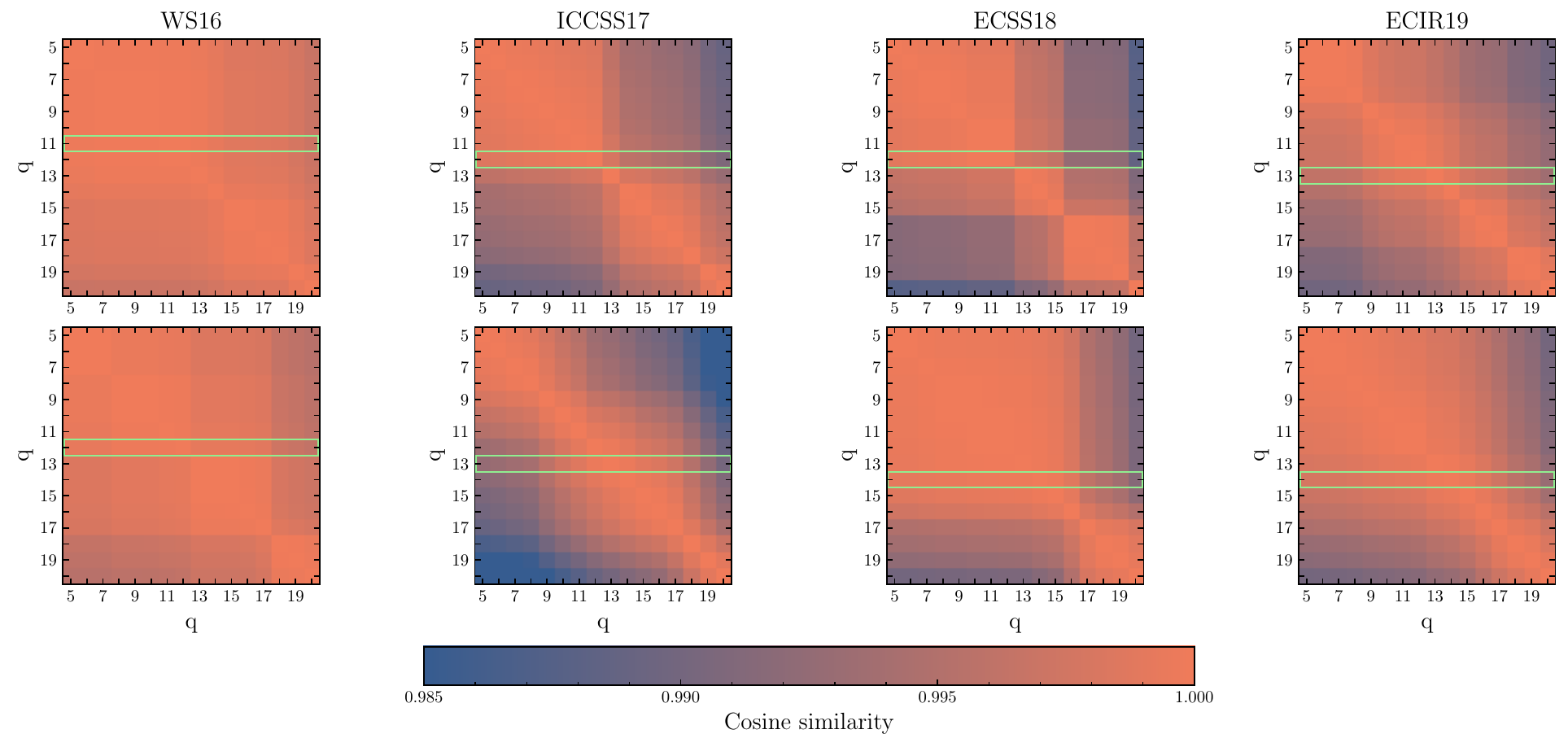}
	\caption{Cosine similarity heat maps over vectors containing all TMDT$_q$ values for $\alpha_{J}$ (top) and $\alpha_{\mathrm{cos}}$ (bottom) in each dataset. Green rectangles represent the $q$ value for which the average similarity is maximum.}
	\label{fig:TMDT_q_rob}
\end{figure}

Finally, because TMDT estimators depends on the percentile $q$, we computed the dispersion for each individual’s $\alpha_{J}$ and $\alpha_{\mathrm{cos}}$ vectors across a range of $q$ values, from the 5th to the 20th percentile. We then compared the resulting TMDT vectors (i.e., one value per node per $q$) using cosine similarity. As underline in Fig.\ref{fig:TMDT_q_rob}, across all conference datasets, we found extremely high similarity scores, with a minimum value around 0.985. This indicates strong robustness of TMDT to variations in $q$. Represented with the green rectangles, the most stable values were consistently obtained for $q$ in the range of 11\% to 14\%. 

\newpage

\section{\label{sec:period}Determination of periods of interest}

\begin{figure}[h!]
	\centering
	\includegraphics[width=1\textwidth]{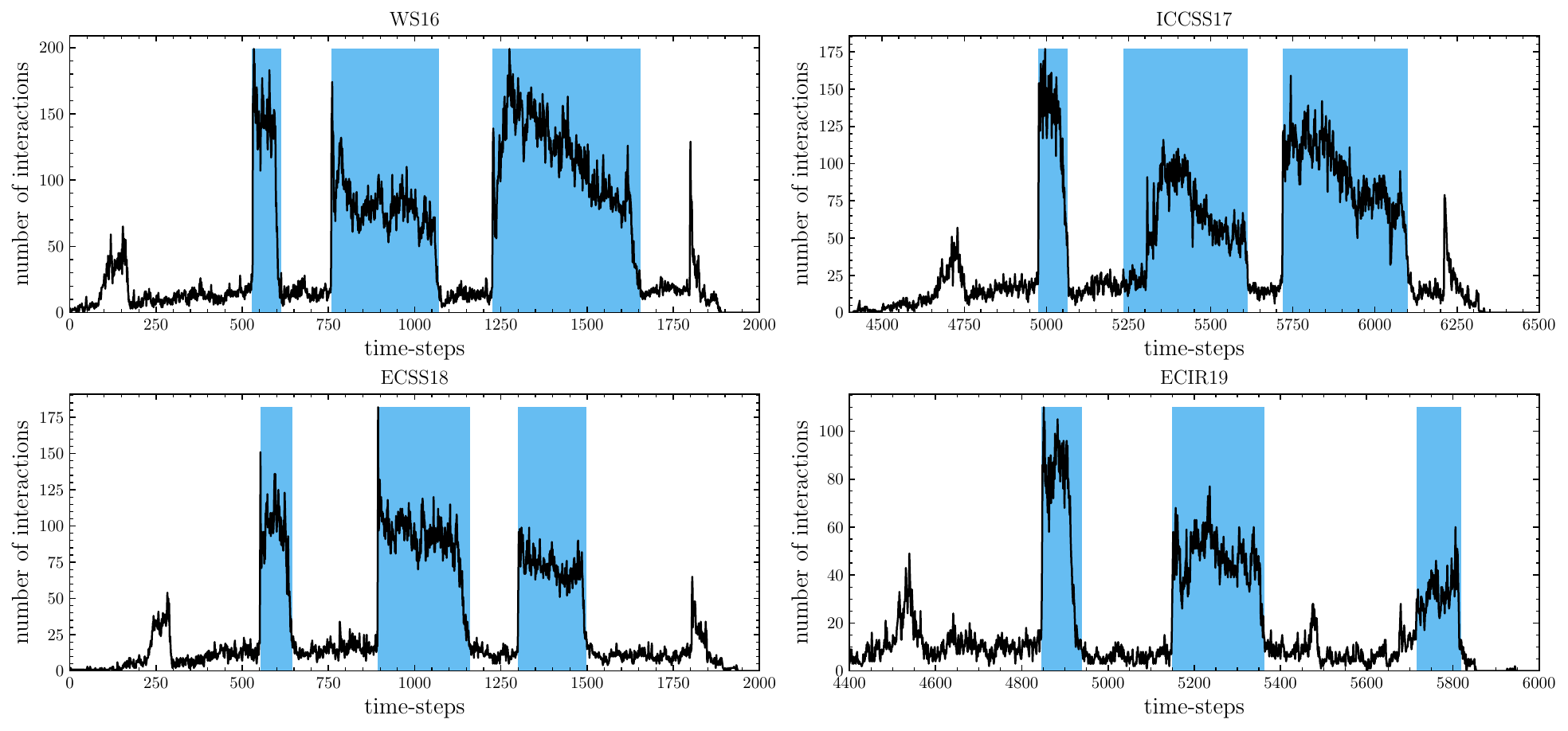}
	\caption{Activity timelines for the first day (WS16 and ECSS18) or the second day (ICCSS17 and ECIR19). Coloured space (in blue) represents the selected time window for aggregation and analysis. The number of time-steps starts at 0 with the first contact register for the dataset.}
	\label{fig:Activity}
\end{figure}

The breaks were identified by examining the activity timelines of each conference, defined as the number of face-to-face interactions recorded at each snapshot. As shown in Fig.~\ref{fig:Activity}, periods of high activity correspond to socially meaningful events such as lunch breaks, coffee breaks, or poster sessions. For each break, we defined the time window as starting at the onset of the activity increase and ending when the activity level returned to its baseline. Small bursts of activity observed at the beginning and end of each day—when attendees arrive or leave the venue—were not included in the analysis, as these transitional periods are less stable and do not reflect the same interaction dynamics as the breaks themselves.

\section{\label{sec:nodes}Number of active nodes}

\begin{table}[h!]
 	\setlength{\tabcolsep}{2pt}
 	\centering
 	\footnotesize
 	\begin{tabular}{l|cc|cc|cc|cc}
 		\toprule
 		 & \multicolumn{2}{c|}{WS16} & \multicolumn{2}{c|}{IC2S2 17} & \multicolumn{2}{c|}{ECSS18} & \multicolumn{2}{c}{ECIR19} \\
 		Minimum contact duration (in $s$) & 0 & 20 &  0 & 20  &  0 & 20  &  0 & 20  \\
 		\midrule
 		
 		Number of nodes active during at least two breaks  & 136 & 136 & 266 & 263 & 157 & 157 & 169 & 168 \\
 		
 		\midrule
 		Number of nodes active at least once & 138 & 138 & 274 & 274 & 164 & 164 & 172 & 172 \\
 		
 		\midrule
 		Proportion lost (\%) & 1.45 & 1.45 & 2.92 & 4.02 & 4.27 & 4.27 & 1.75 & 2.33 \\

 		\bottomrule
 	\end{tabular}
 	\caption{Number and proportion of nodes interacting in at least two breaks compare to the number of node interacting at least once during the conference.}
 	\label{tab:sociodem-agg}
 \end{table}

\newpage
\section{\label{sec:filter}Sensitivity to filtering}
	
\begin{figure}[h!]
	\centering
	\includegraphics[width=1\textwidth]{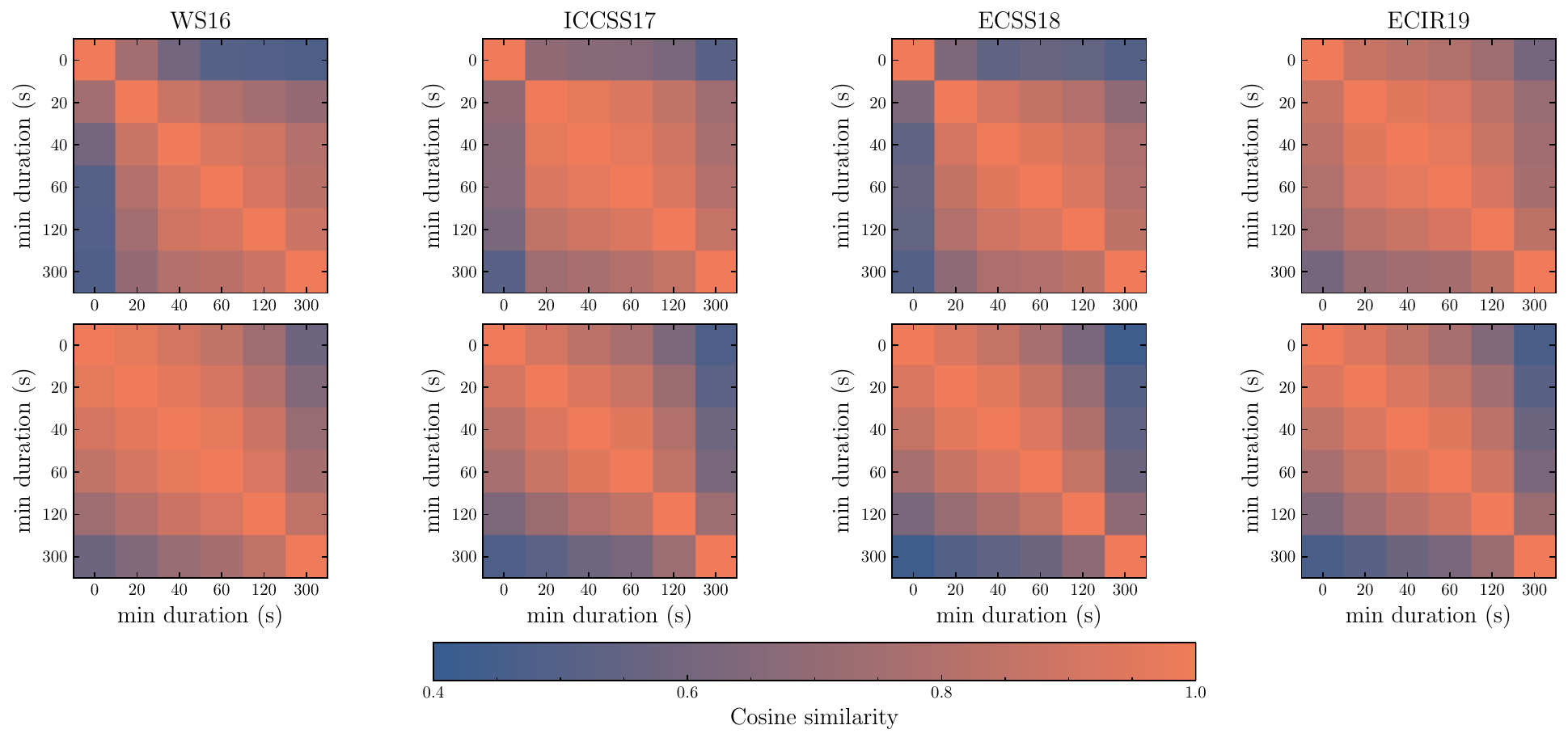}
	\caption{Cosine similarity values applied over the vector containing mean and TMDT$_{12}$ $\alpha_{J}$ (on top) and $\alpha_{\mathrm{cos}}$ (at the bottom) for all nodes in each datasets with removal of contacts with duration less or equal to \textit{min duration}.}
	\label{fig:Robustness}
\end{figure}

Expecting that the removal of contacts based on their duration would affect the $\alpha$ values, we assessed the robustness of our metrics by computing the cosine similarity between the vectors of mean and TMDT$_{12}$ values for all nodes across different filtering thresholds. Specifically, we progressively removed contacts with durations less than or equal to a range of minimum-duration cut-offs, as illustrated in Fig.~\ref{fig:Robustness}. We find that $\alpha_{J}$ is highly sensitive to the removal of the shortest contacts ($20~s$). Such contacts may correspond either to brief but purposeful exchanges between attendees or to incidental, path-crossing events that introduce random behaviour into the data. By contrast, $\alpha_{\mathrm{cos}}$ exhibits much smaller changes under the same filtering, reflecting its greater robustness due to its explicit dependence on interaction weights.

\newpage
\section{\label{sec:rob_filt}Robustness over global dynamic under filtering duration}

\begin{figure}[h!]
	\centering
	\includegraphics[width=1\textwidth]{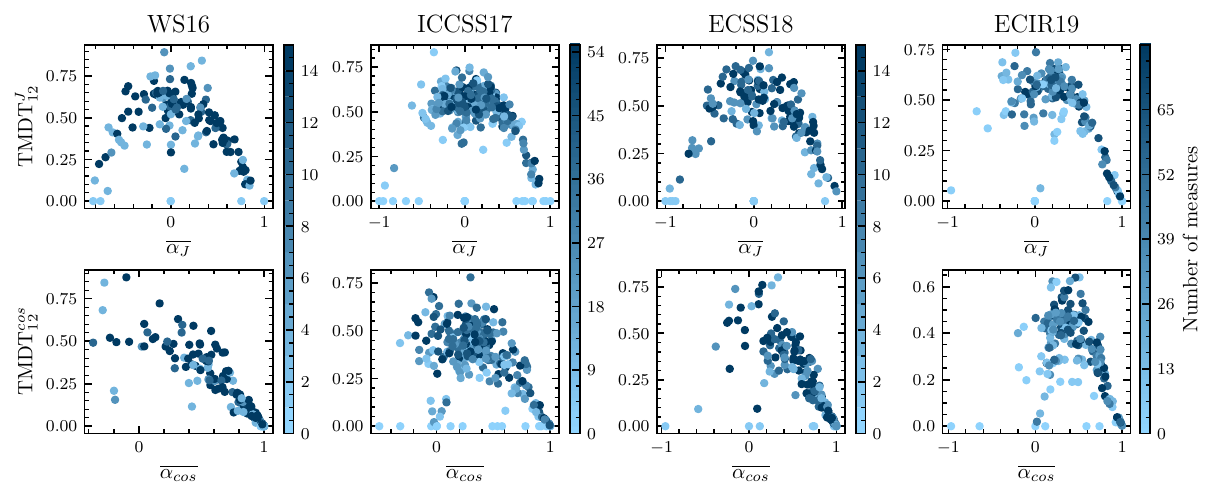}
	\caption{TMDT$_{12}$ versus $\overline{\alpha}$ for each node in each conference without contact removal. For nodes with high dispersion values, $\overline{\alpha}$ is not reliable, whereas for consistent nodes they are meaningful. Point colours indicate the number of measures available for each node (i.e. the number of couple of periods where the node was active) across the full dataset.}
	\label{fig_2_0}
\end{figure}

\begin{figure}[h!]
	\centering
	\includegraphics[width=1\textwidth]{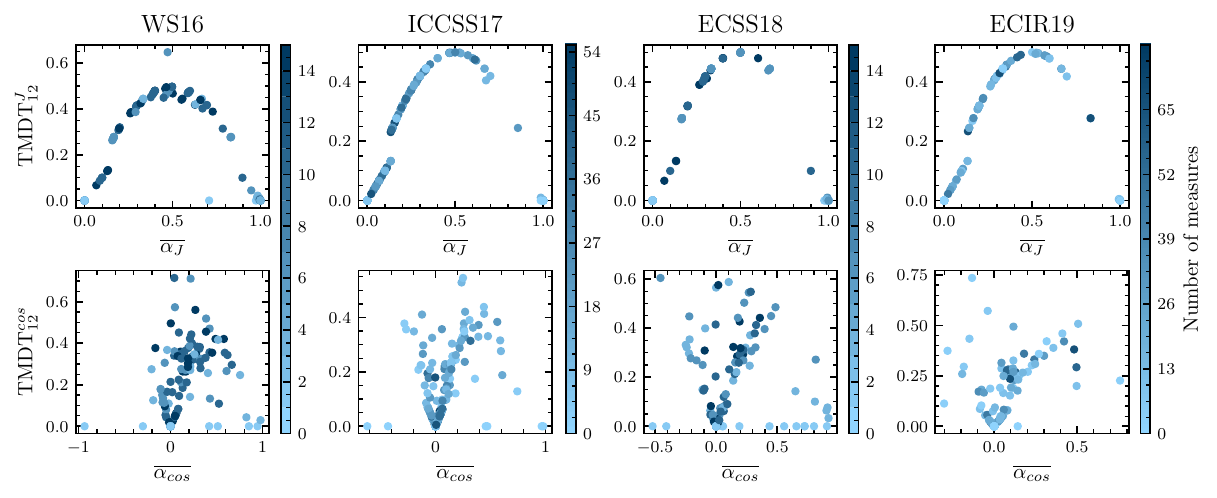}
	\caption{TMDT$_{12}$ versus $\overline{\alpha}$ for each node in each conference with removal of contacts shorter than 300 seconds. For nodes with high dispersion values, $\overline{\alpha}$ is not reliable, whereas for consistent nodes they are meaningful. Point colours indicate the number of measures available for each node (i.e. the number of couple of periods where the node was active) across the full dataset.}
	\label{fig_2_300}
\end{figure}

Looking at Fig.\ref{fig_2_0} and Fig.\ref{fig_2_300}, the reversed U-Shape still holds for the jaccard counterpart with or without the filtering method used. However, this might not be true for the weighted NPC. Comparing both figures, it seems that the removal of contacts less than 300 seconds is also removing nodes from our datasets (either nodes are brighter or disappeared completely). Moreover, Fig.\ref{fig_2_300} shows way less points close to $\overline{\alpha} = 1$ indicating that short contacts play an important role for the persistency of interactions.

\newpage
\section{\label{sec:dyna}NPC dynamic over time}

\begin{figure}[h!]
	\centering
	\includegraphics[width=1\textwidth]{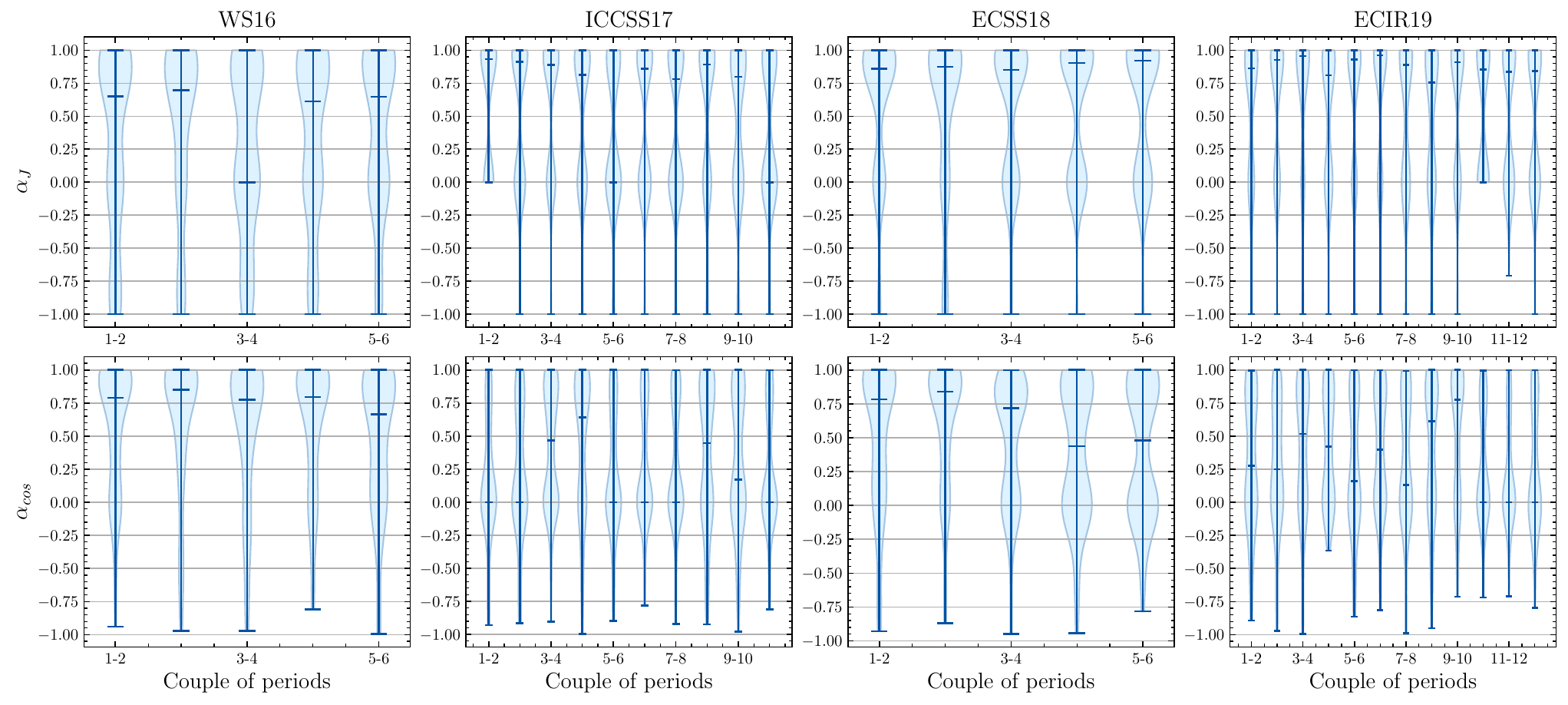}
	\caption{Distributions of $\alpha_{J}$ (top) and $\alpha_{\mathrm{cos}}$ (bottom) for adjacent breaks in each conference dataset. Horizontal bars in between extrema are displaying the median.}
	\label{fig:dynamic}
\end{figure}

From Fig.~\ref{fig:dynamic}, both $\alpha_{J}$ and $\alpha_{\mathrm{cos}}$ show no clear temporal trend. Instead, their values cluster around two distinct modes—one near 0 and one close to 1—indicating that most behaviours are either predominantly random or highly persistent at each period. For WS16 and ECSS18, a gradual decrease in the median $\alpha_{\mathrm{cos}}$ over time suggests that maintaining persistence may become more difficult as the conference progresses. However, this trend is not observed in the longer conferences. Therefore, no generalisable dynamic can be concluded.

\newpage

\section{\label{sec:metadata_corr}Metadata correlation test results}

\begin{table}[h!]
	\setlength{\tabcolsep}{2pt}
	\centering
	\footnotesize
	\begin{tabular}{l|l|lr|lr}
		\toprule
		\multicolumn{2}{c|}{} & \multicolumn{2}{c|}{$\alpha_J$} & \multicolumn{2}{c}{$\alpha_{\mathrm{cos}}$}\\
		\multicolumn{2}{c|}{} & \multicolumn{1}{c}{BM} & \multicolumn{1}{c|}{KS} & \multicolumn{1}{c}{BM} & \multicolumn{1}{c}{KS} \\
		\midrule
		Age& Below 30 (N = 53) &-1.07 (.287) & -0.03 (.771) [.85] & -1.90 (.057) & -0.07 (.033) [.89]\\
		& 30-39 (N = 45) &0.44 (.663) & 0.04 (.736) [.96] & 0.82 (.412) & 0.05 (.264) [.89]\\
		& 40 and more (N = 16) &1.98 (.049) & 0.11 (.089) [-0.55] & \textbf{6.62 (.000)} & \textbf{0.27 (.000) [.47]}\\
		\midrule
		Gender& Male (N = 65) &0.56 (.574) & 0.02 (.965) [.71] & 2.71 (.007) & 0.07 (.020) [.58]\\
		& Female (N = 49) &-0.50 (.614) & -0.02 (.991) [.68] & -2.01 (.045) & -0.05 (.176) [.61]\\
		\midrule
		Country& 1 (N = 63) &2.79 (.005) & 0.06 (.071) [.81] & \textbf{5.00 (.000)} & \textbf{0.11 (.000) [.69]}\\
		& 2 (N = 14) &\textbf{-6.34 (.000)} & \textbf{-0.22 (.000) [.93]} & \textbf{-3.31 (.001)} & \textbf{-0.15 (.001) [.67]}\\
		& Other (N = 33) &0.33 (.739) & 0.05 (.349) [.98] & -2.73 (.007) & -0.08 (.060) [.93]\\
		\midrule
		Language& 1 (N = 47) &2.58 (.010) & 0.07 (.079) [.78] & \textbf{3.44 (.001)} & \textbf{0.12 (.000) [.50]}\\
		& 2 (N = 18) &\textbf{-5.23 (.000)} & \textbf{-0.17 (.000) [.92]} & -2.42 (.016) & -0.12 (.008) [.55]\\
		& Other (N = 42) &0.91 (.361) & 0.05 (.419) [.97] & -0.61 (.541) & -0.03 (.772) [.95]\\
		\midrule
		Academic Status& Student (N = 27) &\textbf{-3.47 (.001)} & \textbf{-0.11 (.005) [.75]} & \textbf{-6.00 (.000)} & \textbf{-0.17 (.000) [.83]}\\
		& PhD (N = 33) &1.96 (.051) & 0.06 (.269) [.60] & 1.43 (.154) & 0.06 (.242) [.64]\\
		& Post-Doc (N = 23) &0.53 (.599) & 0.06 (.399) [.98] & 3.51 (.000) & 0.10 (.013) [.89]\\
		& As. Pr. (N = 8) &1.73 (.086) & 0.11 (.193) [.82] & -0.83 (.407) & -0.10 (.268) [.77]\\
		& Pr. (N = 12) &2.84 (.005) & 0.12 (.080) [.69] & \textbf{4.74 (.000)} & \textbf{0.23 (.000) [.56]}\\
		& Others (N = 4) &-0.61 (.549) & -0.25 (.255) [.96] & -1.02 (.325) & -0.27 (.201) [.95]\\
		\midrule
		Discipline& SHS (N = 49) &-1.11 (.267) & -0.03 (.821) [.52] & -1.62 (.105) & -0.05 (.260) [.85]\\
		& CS (N = 39) &0.93 (.353) & 0.04 (.636) [.87] & \textbf{3.44 (.001)} & \textbf{0.10 (.005) [.90]}\\
		& Maths-Phy (N = 13) &-0.23 (.815) & -0.05 (.880) [.90] & 0.14 (.890) & -0.04 (.989) [.39]\\
		& Other (N = 8) &2.53 (.013) & 0.13 (.076) [-0.53] & -0.67 (.503) & -0.11 (.242) [.98]\\
		\midrule
		Role& Speaker (N = 12) &0.62 (.539) & 0.06 (.844) [.50] & 3.08 (.002) & 0.14 (.027) [.94]\\
		& Poster Presenter (N = 42) &1.39 (.163) & 0.05 (.227) [.92] & -0.82 (.413) & -0.05 (.288) [.98]\\
		& Participant (N = 43) &-0.65 (.516) & -0.05 (.333) [.89] & -0.29 (.769) & -0.02 (1.000) [.92]\\
		& Staff (N = 16) &-1.65 (.100) & -0.08 (.227) [.00] & 1.69 (.092) & 0.10 (.069) [.98]\\
		\bottomrule
	\end{tabular}
	\caption{Brunner–Munzel (BM) and Kolmogorov–Smirnov (KS) tests on WS16. p-values are in parentheses, while the statistic is not. Brackets are used in Kolomogorov-Smirnov to indicates the $\alpha$ value at which the difference in cumulative functions is maximum. Bolds text indicates results considered significant in the article. For anonymisation, countries and languages are coded numerically and not consistent across conferences.}
	\label{tab:sociodem-WS16}
\end{table}

\begin{table}[h!]
	\setlength{\tabcolsep}{2pt}
	\centering
	\footnotesize
	\begin{tabular}{l|l|lr|lr}
		\toprule
		\multicolumn{2}{c|}{} & \multicolumn{2}{c|}{$\alpha_J$} & \multicolumn{2}{c}{$\alpha_{\mathrm{cos}}$}\\
		\multicolumn{2}{c|}{} & \multicolumn{1}{c}{BM} & \multicolumn{1}{c|}{KS} & \multicolumn{1}{c}{BM} & \multicolumn{1}{c}{KS} \\
		\midrule
		Age& Below 30 (N = 57) &\textbf{-3.73 (.000)} & \textbf{-0.05 (.003) [.57]} & \textbf{-3.74 (.000)} & \textbf{-0.06 (.000) [.92]}\\
		& 30-39 (N = 102) &2.16 (.031) & 0.02 (.149) [.95] & \textbf{3.50 (.000)} & \textbf{0.04 (.000) [.92]}\\
		& 40 and more (N = 44) &-1.14 (.253) & -0.03 (.239) [.94] & \textbf{-3.62 (.000)} & \textbf{-0.07 (.001) [.95]}\\
		\midrule
		Gender& Male (N = 138) &0.63 (.532) & 0.01 (.834) [.97] & 0.26 (.799) & 0.01 (.935) [.19]\\
		& Female (N = 65) &-2.27 (.023) & -0.04 (.022) [.52] & -1.76 (.078) & -0.04 (.005) [.19]\\
		\midrule
		Country& 1 (N = 59) &\textbf{-9.66 (.000)} & \textbf{-0.13 (.000) [.89]} & \textbf{-7.32 (.000)} & \textbf{-0.11 (.000) [.11]}\\
		& 2 (N = 42) &\textbf{8.69 (.000)} & \textbf{0.13 (.000) [.52]} & \textbf{8.13 (.000)} & \textbf{0.12 (.000) [.15]}\\
		& 3 (N = 18) &\textbf{4.54 (.000)} & \textbf{0.09 (.000) [.96]} & 2.94 (.003) & 0.05 (.083) [.08]\\
		& 4 (N = 9) &-1.89 (.060) & -0.06 (.296) [.95] & -6.37 (.000) & -0.24 (.000) [.83]\\
		& Other (N = 78) &-0.57 (.567) & -0.02 (.620) [.77] & 0.36 (.719) & 0.01 (.915) [.98]\\
		\midrule
		Language& 1 (N = 52) &\textbf{-4.19 (.000)} & \textbf{-0.07 (.000) [.82]} & \textbf{-5.86 (.000)} & \textbf{-0.10 (.000) [.96]}\\
		& 2 (N = 43) &\textbf{8.31 (.000)} & \textbf{0.12 (.000) [.78]} & \textbf{8.10 (.000)} & \textbf{0.12 (.000) [.01]}\\
		& 3 (N = 22) &\textbf{-8.47 (.000)} & \textbf{-0.17 (.000) [.89]} & \textbf{-8.46 (.000)} & \textbf{-0.18 (.000) [.30]}\\
		& 4 (N = 15) &2.10 (.036) & 0.08 (.007) [.95] & 1.92 (.056) & 0.04 (.482) [.95]\\
		& 5 (N = 9) &-1.89 (.060) & -0.06 (.296) [.95] & -6.37 (.000) & -0.24 (.000) [.83]\\
		& Other (N = 65) &0.07 (.946) & -0.02 (.587) [.98] & \textbf{3.54 (.000)} & \textbf{0.06 (.000) [.82]}\\
		\midrule
		Academic Status& Student (N = 17) &\textbf{-6.41 (.000)} & \textbf{-0.14 (.000) [.85]} & \textbf{-7.35 (.000)} & \textbf{-0.19 (.000) [.63]}\\
		& PhD (N = 65) &0.41 (.682) & 0.02 (.837) [.81] & 0.00 (.996) & 0.02 (.867) [.89]\\
		& Post-Doc (N = 46) &0.39 (.694) & 0.02 (.661) [1.00] & 1.90 (.058) & 0.03 (.162) [-0.14]\\
		& As. Pr (N = 29) &\textbf{3.34 (.001)} & \textbf{0.06 (.002) [.97]} & 1.17 (.243) & 0.03 (.348) [.12]\\
		& Pr. (N = 29) &\textbf{4.22 (.000)} & \textbf{0.09 (.000) [.56]} & \textbf{4.12 (.000)} & \textbf{0.10 (.000) [.72]}\\
		& Other (N = 15) &\textbf{-7.99 (.000)} & \textbf{-0.22 (.000) [.97]} & \textbf{-5.31 (.000)} & \textbf{-0.16 (.000) [.43]}\\
		\midrule
		Discipline& SHS (N = 64) &1.11 (.268) & 0.04 (.026) [.00] & 1.75 (.081) & 0.05 (.000) [.01]\\
		& CS (N = 72) &-2.21 (.027) & -0.04 (.003) [.00] & -0.70 (.485) & -0.04 (.007) [.01]\\
		& Maths-Phy (N = 47) &1.29 (.197) & 0.04 (.094) [.99] & -2.06 (.040) & -0.03 (.327) [.72]\\
		& Other (N = 19) &-2.08 (.038) & -0.05 (.104) [.85] & -1.73 (.083) & -0.05 (.107) [.60]\\
		\midrule
		Role& 1 (N = 92) &\textbf{3.73 (.000)} & \textbf{0.04 (.009) [.96]} & 3.82 (.000) & 0.04 (.012) [.10]\\
		& 2 (N = 44) &\textbf{4.60 (.000)} & \textbf{0.06 (.000) [.89]} & 0.22 (.826) & 0.01 (1.000) [.20]\\
		& 3 (N = 54) &-1.09 (.277) & -0.03 (.419) [.91] & -2.69 (.007) & -0.05 (.005) [.96]\\
		& 4 (N = 16) &\textbf{-19.84 (.000)} & \textbf{-0.36 (.000) [.96]} & \textbf{-8.10 (.000)} & \textbf{-0.24 (.000) [.00]}\\
		\midrule
		Known Persons& None (N = 38) &\textbf{8.75 (.000)} & \textbf{0.13 (.000) [.91]} & \textbf{3.23 (.001)} & \textbf{0.11 (.000) [.01]}\\
		& 1-5 (N = 74) &0.52 (.603) & 0.02 (.503) [.79] & -0.76 (.447) & -0.03 (.087) [.95]\\
		& 6-10 (N = 38) &-1.63 (.104) & -0.04 (.058) [.49] & 0.49 (.623) & 0.04 (.067) [.94]\\
		& 11-20 (N = 28) &\textbf{-3.07 (.002)} & \textbf{-0.07 (.001) [.80]} & 0.49 (.627) & -0.04 (.204) [.07]\\
		& More than 20 (N = 25) &\textbf{-6.71 (.000)} & \textbf{-0.12 (.000) [.91]} & \textbf{-5.60 (.000)} & \textbf{-0.13 (.000) [.00]}\\
		\bottomrule
	\end{tabular}
	\caption{Brunner–Munzel (BM) and Kolmogorov–Smirnov (KS) tests on ICCSS17. p-values are in parentheses, while the statistic is not. Brackets are used in Kolomogorov-Smirnov to indicates the $\alpha$ value at which the difference in cumulative functions is maximum. Bolds text indicates results considered significant in the article. For anonymisation, countries and languages are coded numerically and not consistent across conferences.}
	\label{tab:sociodem-ICCSS17}
\end{table}

\begin{table}[h!]
	\setlength{\tabcolsep}{2pt}
	\centering
	\footnotesize
	\begin{tabular}{l|l|lr|lr}
		\toprule
		\multicolumn{2}{c|}{} & \multicolumn{2}{c|}{$\alpha_J$} & \multicolumn{2}{c}{$\alpha_{\mathrm{cos}}$}\\
		\multicolumn{2}{c|}{} & \multicolumn{1}{c}{BM} & \multicolumn{1}{c|}{KS} & \multicolumn{1}{c}{BM} & \multicolumn{1}{c}{KS} \\
		\midrule
		Age& Below 30 (N = 61) &\textbf{-3.79 (.000)} & \textbf{-0.10 (.001) [.56]} & \textbf{-4.94 (.000)} & \textbf{-0.12 (.000) [.89]}\\
		& 30-39 (N = 64) &1.54 (.125) & 0.04 (.524) [.96] & 2.61 (.009) & 0.08 (.010) [.92]\\
		& 40 and more (N = 14) &1.40 (.164) & 0.11 (.138) [.52] & 1.86 (.065) & 0.15 (.023) [.41]\\
		\midrule
		Gender& Male (N = 76) &-1.08 (.282) & -0.03 (.786) [1.00] & 0.68 (.497) & 0.04 (.446) [.46]\\
		& Female (N = 60) &-0.01 (.991) & -0.04 (.464) [.00] & -2.46 (.014) & -0.08 (.008) [.46]\\
		\midrule
		Language& 1 (N = 51) &\textbf{-4.39 (.000)} & \textbf{-0.09 (.002) [.81]} & -3.21 (.001) & -0.09 (.005) [.90]\\
		& 2 (N = 24) &-0.32 (.752) & -0.05 (.765) [.91] & -1.66 (.098) & -0.08 (.205) [.01]\\
		& Other (N = 55) &1.48 (.140) & 0.04 (.389) [.94] & 1.38 (.168) & 0.04 (.386) [.81]\\
		\midrule
		Academic Status& Student (N = 28) &\textbf{-5.39 (.000)} & \textbf{-0.18 (.000) [.70]} & \textbf{-5.71 (.000)} & \textbf{-0.18 (.000) [.22]}\\
		& PhD (N = 49) &-1.14 (.253) & -0.04 (.522) [.95] & -2.18 (.030) & -0.08 (.019) [.78]\\
		& Post-Doc (N = 35) &1.27 (.204) & 0.04 (.610) [.94] & 1.46 (.146) & 0.05 (.385) [.82]\\
		& As. Pr. (N = 15) &\textbf{5.77 (.000)} & \textbf{0.25 (.000) [.78]} & \textbf{6.60 (.000)} & \textbf{0.30 (.000) [.61]}\\
		& Pr. (N = 2) &-0.76 (.458) & -0.15 (.704) [.92] & 0.94 (.357) & 0.22 (.250) [.41]\\
		& Other (N = 9) &-0.26 (.799) & -0.08 (.659) [1.00] & 2.16 (.033) & 0.18 (.011) [.90]\\
		\midrule
		Discipline& SHS (N = 60) &-0.60 (.550) & -0.03 (.900) [.94] & -1.51 (.131) & -0.04 (.503) [.91]\\
		& CS (N = 48) &-0.46 (.647) & -0.02 (.994) [.88] & 1.53 (.127) & 0.06 (.139) [.92]\\
		& Maths-Phy (N = 8) &2.30 (.024) & 0.15 (.046) [.96] & 1.39 (.168) & 0.11 (.269) [.79]\\
		& Other (N = 22) &-2.29 (.023) & -0.11 (.016) [.71] & \textbf{-3.36 (.001)} & \textbf{-0.13 (.001) [.63]}\\
		\midrule
		Role& Speaker (N = 27) &0.92 (.358) & 0.05 (.745) [.66] & 0.66 (.512) & 0.09 (.073) [.94]\\
		& Poster Presenter (N = 30) &1.00 (.317) & 0.08 (.104) [.98] & -2.21 (.027) & -0.08 (.119) [.66]\\
		& Participant (N = 71) &-2.07 (.038) & -0.05 (.186) [.87] & -2.16 (.031) & -0.06 (.052) [.92]\\
		& Staff (N = 20) &-2.26 (.025) & -0.09 (.058) [.94] & 0.31 (.753) & -0.10 (.038) [.03]\\
		\midrule
		Number of Roles& 1 (N = 130) &-0.54 (.592) & -0.02 (.995) [.66] & -0.23 (.821) & -0.01 (1.000) [.96]\\
		& 2 (N = 9) &-2.07 (.041) & -0.12 (.078) [.91] & \textbf{-4.40 (.000)} & \textbf{-0.22 (.000) [.35]}\\
		\midrule
		Known Persons& None (N = 21) &1.61 (.108) & 0.08 (.245) [1.00] & 2.34 (.020) & 0.13 (.010) [.21]\\
		& 1-5 (N = 66) &0.22 (.825) & 0.02 (.965) [.95] & 0.07 (.945) & 0.03 (.795) [.03]\\
		& 6-10 (N = 25) &-2.92 (.004) & -0.11 (.009) [.82] & -2.05 (.041) & -0.09 (.039) [.53]\\
		& 11-20 (N = 13) &-1.17 (.245) & -0.07 (.585) [.95] & -3.34 (.001) & -0.17 (.001) [.69]\\
		& More than 20 (N = 14) &-1.05 (.295) & -0.05 (.728) [.40] & -0.92 (.358) & -0.10 (.082) [.00]\\
		\bottomrule
	\end{tabular}
	\caption{Brunner–Munzel (BM) and Kolmogorov–Smirnov (KS) tests on ECSS18. p-values are in parentheses, while the statistic is not. Brackets are used in Kolomogorov-Smirnov to indicates the $\alpha$ value at which the difference in cumulative functions is maximum. Bolds text indicates results considered significant in the article. For anonymisation, languages are coded numerically and not consistent across conferences.}
	\label{tab:sociodem-ECSS18}
\end{table}

\begin{table}[h!]
	\setlength{\tabcolsep}{2pt}
	\centering
	\footnotesize
	\begin{tabular}{l|l|lr|lr}
		\toprule
		\multicolumn{2}{c|}{} & \multicolumn{2}{c|}{$\alpha_J$} & \multicolumn{2}{c}{$\alpha_{\mathrm{cos}}$}\\
		\multicolumn{2}{c|}{} & \multicolumn{1}{c}{BM} & \multicolumn{1}{c|}{KS} & \multicolumn{1}{c}{BM} & \multicolumn{1}{c}{KS} \\
		\midrule
		Age& Below 30 (N = 46) &\textbf{-5.01 (.000)} & \textbf{-0.07 (.000) [.00]} & \textbf{-8.75 (.000)} & \textbf{-0.13 (.000) [.77]}\\
		& 30-39 (N = 51) &2.33 (.020) & 0.03 (.055) [.80] & 2.20 (.028) & 0.04 (.012) [.74]\\
		& 40 and more (N = 35) &-1.58 (.113) & -0.03 (.245) [.79] & 2.25 (.025) & 0.04 (.031) [.89]\\
		\midrule
		Gender& Male (N = 106) &-0.18 (.858) & -0.01 (1.000) [.78] & 0.75 (.453) & 0.01 (.892) [.45]\\
		& Female (N = 24) &\textbf{-3.65 (.000)} & \textbf{-0.08 (.000) [.56]} & \textbf{-7.78 (.000)} & \textbf{-0.17 (.000) [.46]}\\
		\midrule
		Language& 1 (N = 30) &-2.61 (.009) & -0.05 (.007) [.98] & 2.17 (.030) & 0.05 (.003) [.94]\\
		& 2 (N = 10) &\textbf{3.19 (.002)} & \textbf{0.10 (.001) [.50]} & \textbf{6.27 (.000)} & \textbf{0.13 (.000) [.00]}\\
		& Other (N = 88) &-2.23 (.026) & -0.04 (.009) [.53] & \textbf{-6.15 (.000)} & \textbf{-0.06 (.000) [.44]}\\
		\midrule
		Academic Status& Student (N = 19) &\textbf{-7.46 (.000)} & \textbf{-0.16 (.000) [.94]} & -2.47 (.014) & -0.10 (.000) [.48]\\
		& PhD (N = 54) &-1.81 (.070) & -0.03 (.151) [.61] & \textbf{-5.24 (.000)} & \textbf{-0.07 (.000) [.86]}\\
		& Post-Doc (N = 12) &\textbf{-3.38 (.001)} & \textbf{-0.08 (.002) [.52]} & \textbf{-3.03 (.003)} & \textbf{-0.09 (.000) [.03]}\\
		& As. Pr. (N = 20) &2.96 (.003) & 0.06 (.014) [.93] & 1.29 (.198) & 0.03 (.676) [.14]\\
		& Pr. (N = 15) &\textbf{5.06 (.000)} & \textbf{0.11 (.000) [.56]} & \textbf{8.32 (.000)} & \textbf{0.17 (.000) [.48]}\\
		& Other (N = 12) &0.74 (.459) & 0.04 (.429) [.93] & 1.43 (.154) & 0.04 (.308) [.96]\\
		\midrule
		Discipline& CS (N = 113) &-0.82 (.410) & -0.02 (.578) [.50] & -1.88 (.060) & -0.02 (.182) [.87]\\
		& Other (N = 19) &\textbf{-3.26 (.001)} & \textbf{-0.10 (.000) [.98]} & -0.83 (.408) & 0.06 (.026) [.93]\\
		\midrule
		Role& Speaker (N = 57) &3.21 (.001) & 0.04 (.007) [.71] & 2.72 (.007) & 0.05 (.002) [.01]\\
		& Poster Presenter (N = 31) &\textbf{5.50 (.000)} & \textbf{0.08 (.000) [.96]} & 0.13 (.894) & 0.03 (.456) [.01]\\
		& Participant (N = 105) &-2.68 (.007) & -0.03 (.032) [.85] & \textbf{-3.77 (.000)} & \textbf{-0.04 (.000) [.23]}\\
		& Staff (N = 21) &\textbf{-23.71 (.000)} & \textbf{-0.32 (.000) [.94]} & \textbf{-10.35 (.000)} & \textbf{-0.27 (.000) [.00]}\\
		\midrule
		Number of Roles& 1 (N = 117) &-0.75 (.452) & -0.01 (.988) [.72] & -1.61 (.108) & -0.02 (.430) [.30]\\
		& 2 (N = 14) &\-3.61 (.000) & -0.09 (.001) [.90] & -1.99 (.047) & -0.06 (.041) [.00]\\
		\midrule
		Known Persons& None (N = 15) &\textbf{16.28 (.000)} & \textbf{0.36 (.000) [.50]} & \textbf{16.12 (.000)} & \textbf{0.35 (.000) [.00]}\\
		& 1-5 (N = 53) &\textbf{-3.18 (.001)} & \textbf{-0.07 (.000) [.00]} & \textbf{-9.60 (.000)} & \textbf{-0.12 (.000) [.77]}\\
		& 6-10 (N = 25) &\textbf{-8.81 (.000)} & \textbf{-0.14 (.000) [.95]} & -1.57 (.117) & -0.08 (.000) [.08]\\
		& 11-20 (N = 16) &\textbf{4.07 (.000)} & \textbf{0.09 (.000) [.68]} & \textbf{3.78 (.000)} & \textbf{0.10 (.000) [.05]}\\
		& More than 20 (N = 23) &-1.69 (.092) & -0.04 (.219) [1.00] & 1.67 (.094) & 0.07 (.001) [.91]\\
		\bottomrule
	\end{tabular}
	\caption{Brunner–Munzel (BM) and Kolmogorov–Smirnov (KS) tests on ECIR19. p-values are in parentheses, while the statistic is not. Brackets are used in Kolomogorov-Smirnov to indicates the $\alpha$ value at which the difference in cumulative functions is maximum. Bolds text indicates results considered significant in the article. For anonymisation, languages are coded numerically and not consistent across conferences.}
	\label{tab:sociodem-ECIR19}
\end{table}